\documentclass[preprint,showpacs,preprintnumbers,amsmath,amssymb,nofootinbib]{revtex4}

\usepackage{graphicx}
\usepackage{dcolumn}
\usepackage{bm}

\begin{document}
\title{\bf Data Combinations Accounting for LISA Spacecraft Motion}

\author{Daniel A. Shaddock}
\email{Daniel.A.Shaddock@jpl.nasa.gov}
\affiliation{Jet Propulsion Laboratory, California Institute of Technology, Pasadena, CA 91109}

\author{Massimo Tinto}
\email{Massimo.Tinto@jpl.nasa.gov}
\altaffiliation [Also at: ]{Space Radiation Laboratory, California
  Institute of Technology, Pasadena, CA 91125}
\affiliation{Jet Propulsion Laboratory, California Institute of Technology, Pasadena, CA 91109}

\author{Frank B. Estabrook}
\email{Frank.B.Estabrook@jpl.nasa.gov}
\affiliation{Jet Propulsion Laboratory, California Institute of Technology, Pasadena, CA 91109}

\author{J.W. Armstrong}
\email{John.W.Armstrong@jpl.nasa.gov}
\affiliation{Jet Propulsion Laboratory, California Institute of Technology,
 Pasadena, CA 91109}

\date{\today} 
\begin{abstract}

LISA is an array of three spacecraft in an approximately equilateral
triangle configuration which will be used as a low-frequency gravitational
wave detector. We present here new generalizations of the Michelson- and Sagnac-type  time-delay interferometry data combinations. These combinations cancel laser phase noise in the presence of different up and down propagation delays in each arm of the array, and
slowly varying systematic motion of the spacecraft.  The gravitational
wave sensitivities of these generalized combinations are the same as previously computed for the stationary cases, although the combinations are now more complicated. We introduce a diagrammatic representation to illustrate that these combinations are actually synthesized equal-arm interferometers.
\end{abstract}

\maketitle


The Laser Interferometer Space Antenna (LISA) \cite{1} is a space-borne
gravitational wave (GW) detector mission which will use coherent laser
beams exchanged between three widely separated spacecraft to study
low-frequency ($10^{-4} - 1$ Hz) GWs.  Modeling each spacecraft as
carrying lasers, beam splitters, photo-detectors and drag-free proof
masses on each of two optical benches, it has been shown \cite{5,6,7}
that the six measured time series of Doppler shifts of the one-way
laser beams between spacecraft pairs, and the six measured shifts
between adjacent optical benches on each spacecraft, can be combined,
with suitable time delays, to cancel the otherwise overwhelming phase
noise of the lasers ($\Delta \nu/\nu \simeq 10^{-13}$) to a strain
level $h \simeq 10^{-23}$.  This technique is called time-delay
interferometry (TDI).  

Initial analyses of TDI \cite{5,6,7,9,9aa}
were for a non-rotating, rigid LISA array. However,the actual LISA orbits \cite{9a,9b} produce annual rotation of the
array.  Recently, TDI applied to a rotating
LISA was considered \cite{10}.  It was shown that the Sagnac effect leads to non-negligible light-time
differences for light traveling around the array in the clockwise and
counterclockwise senses.  New candidate
implementations of TDI involving both one-way measurements and laser phase-locking were presented which obviate the difficulties of a rotating array.
In parallel work, Cornish and Hellings \cite{1a} pointed out that in addition to the rigid body rotation, more general inter-spacecraft velocities can occur (so called ``flexing'' of the LISA constellation). It was shown that this flexing will introduce phase noise at an unacceptable level, given the present laser frequency noise specification.  

In this paper we first generalize the original Michelson TDI
combinations to an array with systematic spacecraft velocities, showing
that these generalizations effectively cancel all laser phase noises
(residuals second order in v/c, much smaller than the secondary LISA
noise levels). We then analyze the generalized Sagnac combinations,
showing that they too cancel laser phase noise adequately in a rotating
and/or shearing LISA array.  Finally, we emphasize that the gravitational
wave sensitivities of these generalized combinations are the same as
those for the non-rotating case.

There are six beams exhanged between the LISA spacecraft, together with
the six phase measurements $s_{ij}$ ($i,j = 1, 2, 3$) recorded when
each transmitted beam is mixed with the laser light of the receiving
optical bench.  The phase fluctuations from the six lasers, which need
to be canceled, can be represented by six random processes $p_{ij}$, 
where $p_{ij}$ is the phase of the laser in spacecraft $j$ on the optical bench
facing spacecraft $i$. In what follows we assume the center frequencies of the lasers are all
the same.

Since the LISA triangular array has systematic motions, the two one-way light
times between any spacecraft pair are not the same \cite{10}.  Delay times for travel between the spacecraft
must now be accounted for depending on the sense of light propagation
along each link when combining these data.  As before, we label the
arms with single numbers given by the opposite spacecraft; e.g., arm 2 (or $2^{'}$) is opposite
spacecraft 2.  We use primed delays to distinguish light-times taken
in the counter-clockwise sense and unprimed delays for the clockwise
light times. (Note that we have changed the data labeling conventions
from that used in all previous papers by the last three authors.  The
subscript labeling of $s_{ij}$ in this paper is that of \cite{10}.)
Explicitly: $s_{23}$ is the one-way phase shift measured at spacecraft
3, coming from spacecraft 2, along arm 1.  The laser phase noise in
$s_{23}$ is $p_{32}(t - L_1) - p_{23}(t)$, where we take c = 1, so
that $L_1$ is the light time in the direction from spacecraft 2 to
spacecraft 3.  Similarly, $s_{32}$ is the phase shift measured on
arrival at spacecraft 2 along arm $1'$ of a signal transmitted from
spacecraft 3.  The laser phase noise in $s_{32}$ is $p_{23}(t -
L_1^{'}) - p_{32}(t)$, where $L_1^{'}$ is the light time in the 
sense from 3 to 2 along arm $1^{'}$.  Due to the relative motion, $L_1
\neq L_1^{'}$ in general.  For the further delays used in the TDI
combinations we use the same conventions, being careful to distinguish
light travel along arms with primes or not, depending on the sense of
the measurement.  For example, our notation for delaying the time
series $s_{32}(t)$ by the clockwise light time in arm 1 would be
$s_{32,1}$ while delaying by the counterclockwise light time in arm $1^{'}$
would be $s_{32,1'}$.  As before, we denote six further data streams,
$\tau_{ij}$ ($i,j = 1, 2, 3$), as the intra-spacecraft
metrology data used to monitor the motion of the two
optical benches and the relative phase fluctuations of the two lasers
on each of the three spacecraft.

Cornish and Hellings \cite{1a} have formulated TDI when the two delay
times on each link, e.g. $L_1$ and $L_1^{'}$ are not only different
(pure rotation) but also themselves functions of time.  In the
subscript notation for delays the  {\it{order}} of the subscripts now
becomes important for laser phase terms.  The subscripts can no longer be permuted freely to
show cancellation of laser noises in the TDI combinations and we will
use a semicolon, instead of a comma, to emphasize this.  (The other,
secondary, noises in LISA are so much smaller, and the rotation and
systematic velocities in LISA are so intrinsically small, that index
permutation may still be done for them.)  
We will then go to first order expansions of the velocity, $\dot{L}$,
dropping quadratic terms in $\dot{L}/c$ and acceleration terms. This iterated time delay method, to first order in the velocity, is illustrated
abstractly as follows.  Given a function of time $\Psi = \Psi(t)$,
time delay by $L_i$ is denoted with the standard comma notation:
\begin{equation}
\Psi_{,i} \equiv \Psi(t - L_i(t))
\label{psi1}
\end{equation}
\noindent
We then impose a second time delay $L_j(t)$:
\begin{eqnarray}
\Psi_{;ij} & \equiv &  \Psi(t - L_j(t) - L_i(t - L_j(t))
\nonumber \\
&  \simeq  & \Psi(t - L_j(t) - L_i(t) + \dot  L_i(t)  L_j) 
\nonumber \\
&  \simeq & \Psi_{,ij} + \dot \Psi_{,ij} \dot L_i L_j
\label{psi2}
\end{eqnarray}

\noindent
A third time delay $L_k(t)$ gives:
\begin{eqnarray}
\Psi_{;ijk} & = & \Psi(t - L_k(t) - L_j(t - L_k(t)) - L_i(t - L_k(t) - L_j(t - L_k(t))))
\nonumber \\
& & \simeq \Psi_{,ijk} + \dot \Psi_{,ijk} [\dot L_i (L_j + L_k) + \dot L_j L_k]
\label{psi3}
\end{eqnarray}
\noindent
and so on, recursively;
each delay generates a correction proportional to its rate of change times
the sum of all delays coming after it in the subscripts.


Consider a constant-length equal-arm interferometer with one-way readouts
for relative phase not only at the (central) spacecraft 1, but also
at outlying spacecraft 2, and 3, with $L_2 = L_2^{'} = L_3 = L_3^{'} = L$.
The phase data $s_{12}$, $s_{21}$, $s_{13}$, $s_{31}$ can be combined to
give the Michelson response
\begin{equation}
S(t) = [s_{31} + s_{13,L}] - [s_{21} + s_{12,L}] \ ,
\label{eq:1}
\end{equation}
\noindent
where we have purposely grouped the terms in square brackets to
indicate that they provide the synthesized two-way phase data from
each arm measured at the central spacecraft \cite{9c}.  With these conditions, the laser phase noise is eliminated in $S$.  An impulsive GW (duration short compared with $L$) will appear four times
in the laser-noise-free time series $S$.

With the {\it{unequal}} arm-lengths of space based
interferometers, the time series $S$ no longer cancels the laser phase
fluctuations, leaving behind a remaining laser noise term that is proportional to the difference in the arm lengths to first order. In
order to correct for this problem, it was shown in \cite{5} that the
two-way measurements entering into $S$ must be differenced, with
suitable delays, to again eliminate laser noise.  This gives the TDI
combination $X$ \cite{5}:
\begin{equation}
X  = [(s_{31} + s_{13,2}) + (s_{21}+ s_{12,3'})_{,22'}] - 
[(s_{21} + s_{12,3'}) + (s_{31} + s_{13,2})_{,33'}] 
\label{eq:2}
\end{equation}

\noindent
together with the analogous time series, $Y$ and $Z$, centered on
spacecraft $2$ and $3$ obtained from equation (\ref{eq:2}) by
permutation of the indices.  The response function of $X$ to a
gravitational wave pulse has twice as many terms as does $S$, and
results in an ``8 pulse'' GW response.

\begin{figure}
\centering
\includegraphics[width=2.7in, angle=0]{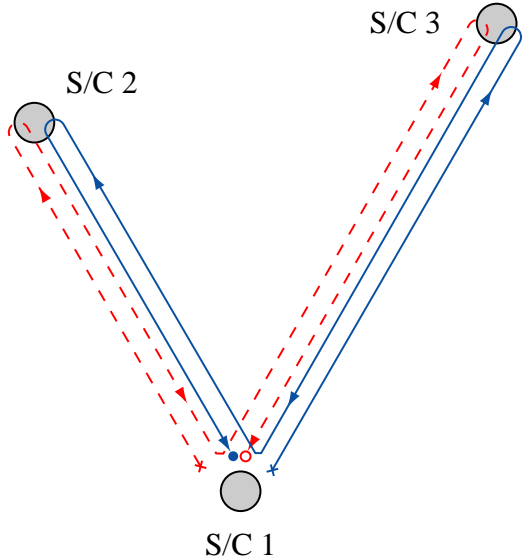}
\caption{Schematic diagram for $X$, showing that it is a synthesized
  zero-area Sagnac Interferometer. The optical path begins at an ``x'' and the measurement is made at an ``o''}
\end{figure}

Equation (\ref{eq:2}) shows that $X$ is the difference of two sums of
phase measurements, each corresponding to a specific light path (the
continuous and dashed lines in Figure 1). The continuous line,
corresponding to the first square-bracket term in equation
(\ref{eq:2}), represents a light-beam transmitted from spacecraft 1 and
made to bounce once at spacecraft 3 and 2 respectively. Since the other
beam (dashed line) experiences the same overall delay as the first
beam (although by bouncing off spacecraft 2 first and then spacecraft 3) when
they are recombined they will cancel the laser phase fluctuations
exactly, having both experienced the same total delays (assuming stationary spacecraft). For this reason the combination $X$ can be regarded as a synthesized (via TDI)
zero-area Sagnac interferometer, with each beam experiencing a delay
equal to $(L_2 + L_{2}' + L_3 + L_{3}')$. In reality there are only two beams in each arm (one in each direction) and the lines in Figure 1 represent the paths of phase information rather than paths of distinct light beams.

If LISA has pure rotation, so that $L_2, L_2^{'}$, etc. are different
but still time-{\it{independent}}, the original TDI combinations can
be easily modified by being careful with the prime-noprime notation.
This has been done in the above equations for X. 
In this case these modified TDI combinations still cancel
all laser noises exactly \cite{1a}. 

In the general case of time-{\it{dependent}} spacecraft separations,
in which systematic velocities enter as a perturbation on the static LISA
unequal arm configuration, it has recently been pointed out by
Cornish and Hellings \cite{1a} that, for the LISA laser stability
specifications of $\simeq 30~$Hz/$\sqrt{\rm Hz}$, the orbital variations,
$\dot L_i$ and $\dot L_i^{'}$, bring in laser phase noise in
the combination $X$ at levels higher than the secondary noises in the
lower part of the LISA frequency band ($10^{-4} \le f < 10^{-3}$ Hz).
The same effect also prevents perfect
cancellation of the phase noise in the other stationary-spacecraft TDI combinations . Improving the laser noise performance by about one order of magnitude, and reducing the LISA arm lengths were suggested as possible solutions to recover the required LISA sensitivity.
Technical challenges may make further frequency stabilization down to this level unfeasible. In the following we show that there is another solution that requires no modification of the LISA hardware: it is possible to generalize TDI combinations ($X, Y, Z$) to remove the
velocity dependence of the laser noise term remaining in the $X$ combination.
This is accomplished by further differencing the synthesized two-way
measurements from each arm according to the diagram shown in Figure 2.
\begin{figure}
\centering
\includegraphics[width=4.0in, angle=0]{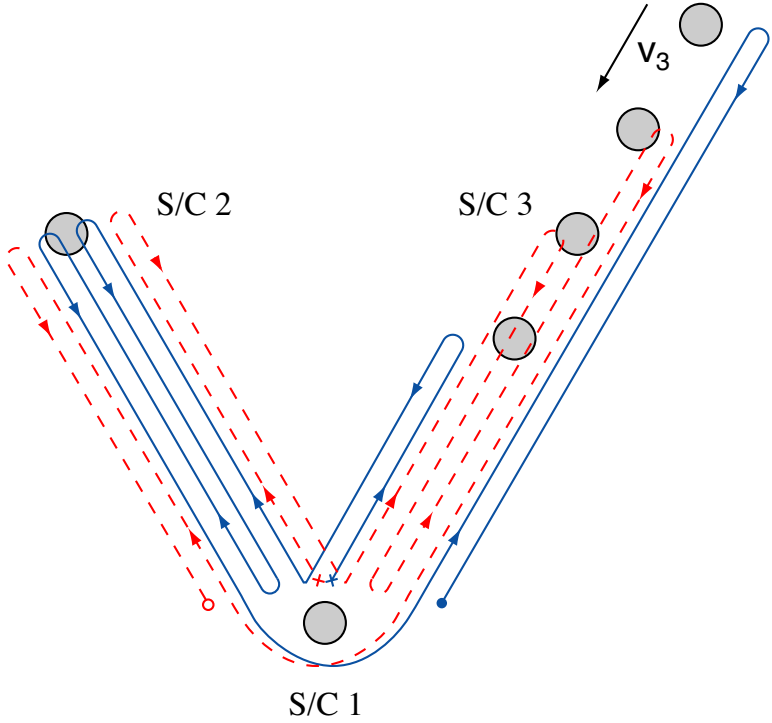}
\caption{Schematic diagram for the new TDI combinations $X_1$. 
  Snapshots of spacecraft are shown at four times.  Note that
  by properly delaying the beams within the two arms it is possible
  to average the effects of constant relative velocities of the
  spacecraft and equalize the optical paths of the two beams (continuous
  and dashed lines).}
\end{figure}
As an example, let us assume that the
velocity of spacecraft $3$ relative to spacecraft $1$ and $2$ is as
shown in Figure 2. In this configuration the effective optical paths
of the two synthesized beams (continuous and dashed lines)
can be described as follows. One of the beams (continuous line, for
instance) is first made to bounce off spacecraft $3$ once, then spacecraft $2$ twice, and finally makes one more bounce off spacecraft $3$ before the phase measurement is made.
Symmetrically, the other beam (dashed line) is first made to bounce
off spacecraft $2$ once, then spacecraft $3$ twice, and
once again off spacecraft $2$ before the phase measurement is made. By delaying the beams in this manner we are able to average out the changes of the arm
lengths taking place over a round-trip-light-time, making the two
optical paths of the two beams essentially equal. 

The diagram shown in Figure 2 can be converted into a specific linear
combination of the inter-spacecraft one-way phase measurements,
$s_{ij}$, and metrology measurements performed onboard each spacecraft
$\tau_{ij}$. This new TDI combination, $X_1$, is very insensitive to 
``flexing'', and is given by:

\begin{eqnarray}
X_1 & = & [(s_{31} + s_{13;2}) + (s_{21} + s_{12;3'})_{;2'2}
+ (s_{21} + s_{12;3'})_{;33'2'2}
+ (s_{31} + s_{13;2})_{;33'33'2'2}]
\nonumber
\\
& & -
[(s_{21} + s_{12;3'})
+ (s_{31} + s_{13;2})_{;33'}
+ (s_{31} + s_{13;2})_{2'233'} +
(s_{21} + s_{12;3'})_{;2'22'233'}]
\nonumber
\\
& & +  {1 \over 2} \ [(\tau_{21} - \tau_{31}) - (\tau_{21} -
\tau_{31})_{;33'} - (\tau_{21} - \tau_{31})_{;2'2}
+ (\tau_{21} - \tau_{31})_{;33'33'2'2}
\nonumber
\\
& & + (\tau_{21} - \tau_{31})_{;2'22'233'}
- (\tau_{21} - \tau_{31})_{;2'233'33'2'2}]
\label{eq:3}
\end{eqnarray}

\noindent
(with $X_2$ and $X_3$ obtained by cyclic permutation of the spacecraft
indices.)  Substituting into equation (\ref{eq:3}) the laser phase noise terms entering the $s_{ij}$ and $\tau_{ij}$, and applying the
expansion rules of equations (\ref{psi1} - \ref{psi3}), it can be shown
that, to first order in the systematic relative velocities of the
spacecraft, laser phase noise is once again eliminated.  This degree of
suppression puts the laser noises several orders of
magnitude below the secondary noises, even for a rotating/shearing
LISA array.  Ultra-stable oscillator (USO) noises, which enter in the
phase measurements of $X_1$, can be calibrated and removed in the same
manner as for $X$ \cite{9}.  

Note that if one were to implement the locking configuration
described in \cite{11}, in which one of the lasers in spacecraft $1$
is the master and the remaining five are slaved to it (so
$s_{12}=s_{13} = \tau_{21} - \tau_{31} = \tau_{32} - \tau_{12} = \tau_{13} - \tau_{23} =
0$), the expression above for $X_1$ would reduce to

\begin{eqnarray}
X_{1locked} & = & [s_{31} + s_{21;2'2} + s_{21;33'2'2} + s_{31;33'33'2'2}]
\nonumber
\\
& & - [s_{21} + s_{31;33'} + s_{31;2'233'} + s_{21;2'22'233'}] \ ,
\label{eq:4}
\end{eqnarray}
\noindent
where the two data sets $s_{31}$, $s_{21}$ should now be regarded as
two-way measurements.


As with $X_1$, the solution to ``flexing'' for
the Sagnac combination $\alpha$ can be obtained by analyzing the diagram shown in
Figure 3. Consider the sequence of one-way measurements
starting from spacecraft $1$, and let us propagate counterclockwise and
clockwise the (synthesized) beams, represented by the continuous and
dashed lines in Figure 3. 
\begin{figure}
\centering
\includegraphics[width=3.5in, angle=0]{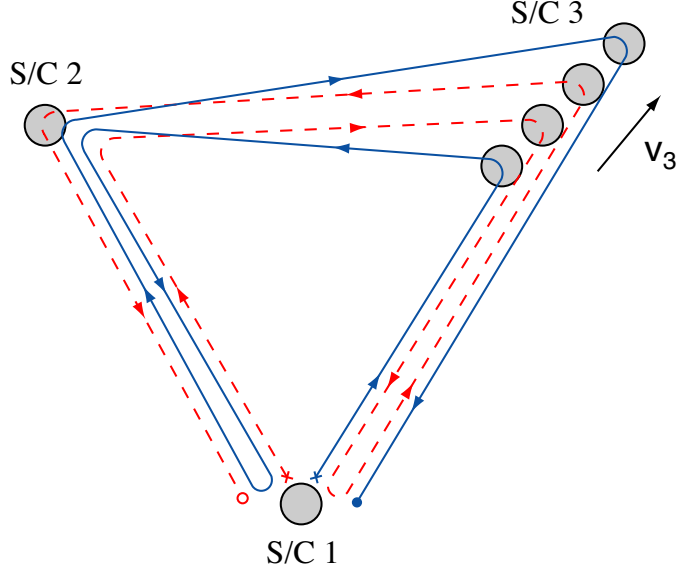}
\caption{ Schematic diagram for the generalized TDI combination $\alpha_1$. 
  By making each of the two beams circulate clockwise and counterclockwise
  once, it is possible to make their optical paths (continuous and dashed lines) essentially equal. (See text.)}
\end{figure}
Once again, for this demonstration we have assumed spacecraft $3$ to have a
non-zero velocity relative to the other two (stationary) spacecraft.
The continuous and dashed lines correspond to the synthesized optical
paths of the light transmitted by spacecraft $1$.  As Figure 3
suggests, this way of combining the one-way measurements compensates for the arm length imbalance generated by the motion of spacecraft $3$ during the round-trip time, making the delays experienced by the two synthesized beams essentially equal. By reference to
Figure 3, the expression for
the generalized Sagnac observable $\alpha_1$ is:
\begin{eqnarray}
\alpha_1 & = & [s_{31} + s_{23;2} + s_{12;12} +
s_{21;312} + s_{32;3'312} + s_{13;1'3'312}] 
\nonumber \\
& & - [s_{21} + s_{32;3'} + s_{13;1'3'} +
s_{31;2'1'3'} + s_{23;22'1'3'} + s_{12;122'1'3'}]
\nonumber
\\
& & + {1 \over 2} [(\tau_{21} - \tau_{31}) - (\tau_{21} -
\tau_{31})_{;2'1'3'312}
+ (\tau_{32} - \tau_{12})_{;3'}
\nonumber \\
& & + (\tau_{32} - \tau_{12})_{;12} - (\tau_{32} - \tau_{12})_{;3'312}
- (\tau_{32} - \tau_{12})_{;122'1'3'}
\nonumber \\
& & + (\tau_{13} - \tau_{23})_{;2} + (\tau_{13} - \tau_{23})_{;1'3'}
- (\tau_{13} - \tau_{23})_{;22'1'3'}
\nonumber\\
& & - (\tau_{23} - \tau_{13})_{;1'3'312}]
\label{eq:5}
\end{eqnarray}
This expression coincides with that first derived in
\cite{10} except that
now the order in which the delays are applied to the one-way
measurements is important (as shown by the presence of a semicolon in
equation (\ref{eq:5})).
\noindent
Expanding this expression using the results of equations (\ref{psi1} - \ref{psi3}) we
find that, unlike $X_1$, the residual phase noise of the laser(s) at
spacecraft 1 remains to first order in the systematic
velocities.  The residual phase noise in $\alpha_{1}$ is:

\begin{eqnarray}
\dot p_{21,1231'2'3'} [(\dot L_1^{'} + \dot L_2^{'} +
\dot L_3^{'}) (L_1 + L_2 + L_3)
  - (\dot L_1 + \dot L_2 + \dot L_3) (L_1^{'} + L_2^{'} + L_3^{'})]
\end{eqnarray}

\noindent
Fortunately, although first order in the relative velocities, the residual is small, as it involves the difference of the clockwise and counterclockwise rates of change of the propagation delays on the {\it{same}} circuit.  For LISA, the remaining laser phase noises
in $\alpha_i$,  $i = 1, 2, 3$, are several orders of magnitude
below the secondary noises.  LISA's GW sensitivity in
$\alpha_1$ is essentially the same as for $\alpha$.


Since these generalized Michelson- and Sagnac-type TDI combinations involve twice as many terms as
those entering into the original (stationary array) and modified
(rotating array) TDI observables, the requirements in accuracies and
precisions in physical quantities needed to synthesize the TDI
combinations (such as arm length knowledge, synchronization of the
onboard clocks, etc.) change.  Our preliminary analysis indicates that
the arm length and onboard clocks synchronization
accuracies (needed to suppress laser noise to below secondary noise
sources in these new TDI combinations) will be more stringent than those
previously estimated for the original TDI combinations by a factor of
about $2$. Further analysis is needed on these requirements and on practical
issues concerning the implementation of TDI for LISA.

This research was performed at the Jet Propulsion Laboratory,
California Institute of Technology, under contract with the National
Aeronautics and Space Administration.


\begin{references}
\bibitem{1} P.L. Bender, K.  Danzmann,\& the LISA Study Team,
{\it{Laser \ Interferometer \ Space \ Antenna \ for \ the \ Detection \ of
\ Gravitational \ Waves, \ Pre-Phase \ A \ Report}},
$\bf{MPQ 233}$ (Max-Planck-Instit\"ut f\"ur
Quantenoptik, Garching), July 1998.
\bibitem{5} M. Tinto \& J.W. Armstrong, {\it Phys. Rev. D}, {\bf 59}, 102003 (1999).
\bibitem{6} J.W. Armstrong, F.B. Estabrook \& M. Tinto, {\it Ap. J.}, {\bf 527}, 814 (1999)
\bibitem{7} F.B. Estabrook, M. Tinto \& J.W. Armstrong, {\it Phys. Rev. D}, {\bf 62}, 042002 (2000)
\bibitem{9} M. Tinto, F.B. Estabrook \& J.W. Armstrong {\it Phys. Rev. D}, {\bf 65}, 082003 (2002)
\bibitem{9aa} S. V.Dhurandhar, K. R. Nayak and J.-Y. Vinet {\it Phys. Rev. D}, {\bf 65}, 102002 (2002)
\bibitem{9a} M.A. Vincent \& P.L. Bender, {\it Proc. Astrodynamics Specialist Conference
    (Kalispell, Montana)} {\bf 1} (San Diego: Univelt) 1346 (1987).
\bibitem{9b} W.M. Folkner, F. Hechler, T.H. Sweetser, M.A.  Vincent 
   \&  P.L. Bender {\it Class. Quantum Grav.}, {\bf 14}, 1405 (1997).
\bibitem{10} D. A. Shaddock gr-qc/0306125v1 (2003), submitted to {\it Phys. Rev. D.}
\bibitem{1a} N. J. Cornish \& R. W. Hellings gr-qc/0306096v2 (2003)
\bibitem{9c} M. Tinto, {\it Phys. Rev. D}, {\bf 58}, 102001 (1998)
\bibitem{11} M. Tinto, D.A. Shaddock, J. Sylvestre, \& J.W. Armstrong,
  {\it Phys. Rev. D}, {\bf{67}}, 122003 (2003)
\end{references}
\end{document}